\documentclass[12pt]{article}
\usepackage[utf8]{inputenc}
\usepackage{geometry}
\usepackage{graphicx}
\usepackage{array}
\usepackage{subfig}
\usepackage{slashed}
\usepackage{amsmath}
\usepackage{amsfonts}
\usepackage{amssymb}
\usepackage{tensor}
\usepackage[cm]{fullpage}
\usepackage{cite}
\usepackage{epstopdf}
\usepackage{comment}
\usepackage{hyperref}
\usepackage{cancel}
\usepackage{mathrsfs}

\usepackage{cleveref}
\crefname{section}{§}{§§}
\Crefname{section}{§}{§§}

\usepackage{booktabs} 

\numberwithin{equation}{section}

\def\0{{(0)}}
\def\1{{(1)}}
\def\2{{(2)}}

\def\<{\langle }
\def\>{\rangle }

\newcommand{\bea}{\begin{eqnarray}}

\newcommand{\eea}{\end{eqnarray}}

\newcommand{\be}{\begin{equation}}
\newcommand{\ee}{\end{equation}}
\newcommand{\ba}{\begin{align}}
\newcommand{\ea}{\end{align}}

   \makeatletter
  \let\over=\@@over \let\overwithdelims=\@@overwithdelims
  \let\atop=\@@atop \let\atopwithdelims=\@@atopwithdelims
  \let\above=\@@above \let\abovewithdelims=\@@abovewithdelims
\renewcommand\section{\@startsection {section}{1}{\z@}%
                                   {-3.5ex \@plus -1ex \@minus -.2ex}
                                   {2.3ex \@plus.2ex}%
                                   {\normalfont\large\bfseries}}

\renewcommand\subsection{\@startsection{subsection}{2}{\z@}%
                                     {-3.25ex\@plus -1ex \@minus -.2ex}%
                                     {1.5ex \@plus .2ex}%
                                     {\normalfont\bfseries}}

\newcommand{\beq}{\begin{equation}}
\newcommand{\eeq}{\end{equation}}
\newcommand{\beqa}{\begin{eqnarray}}
\newcommand{\eeqa}{\end{eqnarray}}
\newcommand{\beqar}{\begin{eqnarray*}}

\def\[{\[}
\def\]{\]}

\newcommand{\bd}[1]{\begin{fmffile}{#1}\begin{fmfgraph*}}
\newcommand{\ed}{\end{fmfgraph*}\end{fmffile}}

\begin{document}

\begin{titlepage}

\begin{flushright}
{\small  CERN--TH--2018--144},
{\small  MPP--2018--142},
{\small LMU--ASC 39/18}
\end{flushright}

\vspace{0.6cm}

\begin{center}

{\LARGE{\textsc{Aspects of  Weyl Supergravity}}}

\vskip0.5cm




\vspace{1cm}
{\large \bf Sergio Ferrara$^{a,b,c}$,   \, Alex Kehagias$^{d}$,\, Dieter L\"ust$^{a,e,f}$}

\vspace{1cm}

{\it

\vskip-0.3cm
\centerline{ $^{\textrm{a}}$ CERN, Theory Department,}
\centerline{1211 Geneva 23, Switzerland}
\medskip
\centerline{ $^{\textrm{b}}$ INFN, Laboratori Nazionali di Frascati,}
\centerline{Via Enrico Fermi 40, 00044 Frascati, Italy}
\medskip
\centerline{ $^{\textrm{c}}$ Department of Physics and Astronomy}
\centerline{and Mani L. Bhaumik Institute for Theoretical Physics, U.C.L.A}
\centerline{Los Angeles CA 90095-1547, U.S.A.}
\medskip
\centerline{ $^{\textrm{d}}$ Physics Division, National Technical University of Athens}
\centerline{ 15780 Zografou Campus, Athens, Greece}
\medskip
\centerline{ $^{\textrm{e}}$ Arnold--Sommerfeld--Center for Theoretical Physics,}
\centerline{Ludwig--Maximilians--Universit\"at, 80333 M\"unchen, Germany}
\medskip
\centerline{$^{\textrm{f}}$ Max--Planck--Institut f\"ur Physik,
Werner--Heisenberg--Institut,}
\centerline{ 80805 M\"unchen, Germany}

}

\end{center}

\vskip0.5cm
\abstract{{
In this paper we study the spectrum of all conformal, ${\cal N}$-extended supergravities (${\cal N}=1,2,3,4$) in four space-time dimensions. When these theories are obtained as massless limit of Einstein 
plus Weyl$^2$
 supergravity, the appropriate
counting of the enhanced gauge symmetries allow us to 
derive the massless spectrum which consist of a dipole ghost graviton multiplet, a ${\cal N}$-fold tripole ghost gravitino, the third state belonging to a spin 3/2 multiplet and a residual vector multiplet 
present for non-maximal ${\cal N}<4$ theories.
These theories are  not expected to have a standard gravity holographic dual in five dimensions.
}}

\end{titlepage}

\newpage

\tableofcontents
\break

\section{Introduction}

There is recently a renewed interest in higher curvature theories. These are theories of the general form $R+{\cal  R}^n$, where ${R}$ is the standard Einstein term and ${\cal R}^n$ denotes collectively $n^{\rm th}$ power of the Riemann, Ricci, Weyl tensors or the curvature scalar \cite{Stelle,Boulware,David, Horowitz,Deser,tHooft,Maldacena,LPS}. A recent discussion on this class of models can be found in  \cite{Alvarez-Gaume:2015rwa}.
In particular, the $R + R^2$ supergravity has  been studied
\cite{Cecotti}, especially in connection to the inflationary dynamics\cite{staro}. 
%
In fact, the  $R+R^2$ theory, known as the Starobinsky model \cite{Starobinsky} for inflation, propagates besides the usual massless graviton, an additional massive spin-0 state,  known as the ``scalaron field"  or  the so called ``no-scale field". 
It is this mode that can be  identified with the inflaton field and makes the theory so appealing as  inflation is driven entirely by gravity itself and not by some external scalar field. 
Furthermore, the $R+R^2$ theory can also be embedded consistently in supergravity, whereas the linearized ${\cal N}=1$ theory has been analysed in \cite{Ferrara:1978rk} and the ${\cal N}=2$   in \cite{deWit:1979dzm}. 

On the other hand the $R+R^2$ theory does not include all possible quadratic curvature theories. Indeed,  a second independent quadratic curvature invariant is the square of the Weyl tensor, whereas terms quadratic in the Riemann (or Ricci) tensor can be traded for a Weyl square term and the 4D  Gauss-Bonnet  topological term. 
However, when the Weyl square term is included in the low-energy gravitational effective action, the spectrum changes and includes an additional massive spin-2 ghost \cite{Stelle}.\footnote{It
should be noted that the problem with such ghosts states is that one cannot maintain 
at the same time unitarity and forward propagation in time of positive energy states. Indeed with the opposite $+i\epsilon$ choice one propagates negative energies forward in time but unitarity and the optical theorem is preserved, whereas with the usual $-i\epsilon$ prescription,  ghosts carry positive energy but negative norm \cite{cline}. } 

 In the present paper we first discuss 
the massive Weyl$^2$  theory and its supersymmetric extensions, namely ${\cal N}$-extended Weyl$^2$ supergravities,
which contain not only the 
Weyl${}^2$ term but also the Einstein term $m^2 R$.\footnote{A very interesting  double copy construction of Weyl${}^2$ (super) gravities was recently provided 
in   \cite{Johansson:2018ues}.}
The latter can therefore be seen as  a mass deformation
of the massless theory. 
In addition, due to the relation between massive and massless Weyl supergravity, the
bound ${\cal N}=8$ of Poincare supergravity tranfers to ${\cal N}=4$ 
in the case of
Weyl supergravity \cite{deWit:1978pd}.

Second, we are particularly interested in 
the massless limit $m\to 0$ of the $m^2 R+\mbox{Weyl}^2$ theory 
%
in conformal supergravity \cite{FK1,K2,FvP}. Although these theories contain propagating ghosts, they are nevertheless very interesting, since in
the massless limits, i.e. in the absence of the Einstein term, they provide unique examples of (super)conformal gravitational theories with up to four derivative terms.
Namely in the limit $m\to 0$ the spectrum gets 
re-organized  and the symmetry gets enhanced, namely
from 
(super) Poincare  is enhanced to (super) conformal. In additional also the R-symmetries become local gauge symmetries.
Furthermore in this limit, there are various primary operators, like the Weyl tensor itself.
Another  conformal tensor is the Bel-Robinson tensor, which is basically the square of the Weyl tensor.
 They have a well-defined conformal weight, i.e. transform under conformal transformations in a homogeneous way.
This is true in the massive case and also in the limit of zero mass, i.e. the limit is continuous with respect to the scaling weights. 
Furthermore the Bel-Robinson tensor is conserved in the massive theory on Ricci-flat spaces.

The paper is organized as follows: In section two we describe the bosonic 
Weyl$^2$ gravity, its spectrum and its higher dimensional operators. In section three we discuss the spetcrum of the  super-Weyl theory. Here we again provide some
details of the higher dimensional operators using ${\cal N}=1$ superfield language.
In sections  four, five and six,  the spectra of the  ${\cal N}=2,~{\cal N}=3$ and ${\cal N}=4$ super-Weyl$^2$ theories, respectively, are determined.
We close in section seven with some discussions and expectations on the holographic duals of the (super)conformal Weyl$^2$ gravity.


\section{Bosonic Weyl gravity}

\subsection{Massive theory}

Let us first recall the bosonic Einstein plus (Weyl)$^2$ gravity theory in four dimensions. 
More details can be e.g. found in \cite{Alvarez-Gaume:2015rwa,Salvio:2018crh}.
The action up to four orders in derivatives  has the following form:
\begin{eqnarray}
S
=\int d^4 x\sqrt{-g}\Big{(}{a\over 2}  W_{\mu\nu\rho\sigma}W^{\mu\nu\rho\sigma}+\kappa^2 R\Big{)}
. \label{W4}
\end{eqnarray}
Here $W_{\mu\nu\rho\sigma}=R_{\mu\nu\rho\sigma}-
g_{\mu[\sigma}R_{\rho]\nu}-g_{\mu[\rho}R_{\sigma]\nu}-R/3 g_{\mu[\rho}g_{\sigma]\nu}$ is the Weyl tensor.
The Weyl$^2$-term in the action possesses conformal invariance as it is invariant under the conformal transformation
\begin{eqnarray}
g_{\mu\nu}\to \widehat{g}_{\mu\nu}=\Omega^2 g_{\mu\nu}, \label{conf}
\end{eqnarray}
which leaves the Weyl tensor inert
\begin{eqnarray}
\widehat{W}^\mu_{~\nu\rho\sigma}= W^\mu_{~\nu\rho\sigma} .
\end{eqnarray}
However the Einstein-term is not invariant under conformal transformations, since 
 $R$ transforms under conformal trasformations as:
 \begin{eqnarray}
\widehat{R}= \Omega^{-2}R - 6 \Omega^{-3} g^{\mu\nu}\nabla_\mu 
\nabla_\nu \Omega.  
\end{eqnarray}
Therefore the Einstein-term can be regarded as the mass term in this theory, i.e. a mass deformation, which explicitly breaks conformal invariance.

The equations of motion which follow from (\ref{W4})  are written as 
\begin{eqnarray}\label{weq}
  B_{\mu\nu}+{2\kappa^2\over a} G_{\mu\nu}=0,
 \end{eqnarray} 
 where   $B_{\mu\nu}$ is the Bach tensor  
\begin{eqnarray}
 B_{\mu\nu}=\nabla^\rho\nabla_\sigma W^\sigma_{~\mu\rho\nu}+
 \frac{1}{2}R^{\rho\sigma} W_{\rho\mu\sigma\nu},   \label{bach}
 \end{eqnarray} 
with
\begin{eqnarray}
\widehat B_{\mu\nu}=\Omega^{-2} B_{\mu\nu}. 
\end{eqnarray}
 Note that the second term in $B_{\mu\nu}$ is needed such that the Bach tensor transforms with a uniform weight under conformal transformations.
 The Bach tensor is symmetric, traceless due to conformal invariance and divergence-free (due to diff. invariance)
 \begin{eqnarray}
  B^\mu_{~\mu}=0, ~~~\nabla^\mu B_{\mu\nu}=0, 
  \end{eqnarray} 
and $G_{\mu\nu}$ is the Einstein tensor.

Now we can recall   the propagating modes corresponding to this action. For this, one analyzes  the poles in the propagators 
generated by its quadratic part. 
Specifically, there are two kinds of propagating modes \cite{Stelle}:

\vskip0.5cm
\noindent
(i) A  massless helicity-$\pm2$ graviton $g_{\mu\nu}$. This mode is independent of the couplings $a$ and $\kappa^2$ and it is the standard massless spin-two graviton.

\vskip0.2cm
\noindent
(ii)  A massive spin-two particle $w_{\mu\nu}$ with mass $\kappa^2/a$. It is related to the Weyl$^2$ term in the action. In fact, this massive spin two particle is a ghost for $a>0$, destroying
unitarity, or a tachyon for $a<0$, leading to an instability.  We will call this part of the spectrum the non-standard sector of the theory.
\vskip0.2cm
\noindent
Hence in summary, the Einstein plus (Weyl)$^2$ gravity theory contains seven propagating degrees of freedom.

\subsection{Massless theory}

In the following we consider  the massless limit $\kappa=0$, which is  a pure  Weyl$^2$ theory with  action
\begin{eqnarray}
S_{{\rm  Weyl}^2}={1\over 2 g^2} \int d^4 x\sqrt{-g} W_{\mu\nu\rho\sigma}W^{\mu\nu\rho\sigma}, \label{W2}
\end{eqnarray}
where $g^2=1/a$ is a dimensionless coupling.
Now 
the equations of motion are simply
\begin{eqnarray}\label{weq}
 B_{\mu\nu}=0.
 \end{eqnarray} 
 At the linearized level of the theory, the conformally invariant Weyl equation of motion simply looks like \cite{FZ}
  \begin{equation}\label{linWeyl}
\partial^\mu\partial^\rho W_{\mu\nu\rho\sigma}=0\, .
\end{equation}

\vskip0.5cm\noindent
The pure Weyl$^2$ possesses conformal invariance and it propagates six degrees of freedom \cite{Rieg}:
 \vskip0.3cm
\noindent
(i) The standard  massless spin-two graviton, corresponding to a planar wave in Einstein gravity.

\vskip0.2cm
\noindent
(ii)  In the non-standard sector there is massless spin-two ghost particle, which corresponds to a non-planar wave.  In addition there is a massless vector, which originates from the 
$\pm 1$ helicities of the massive $w_{\mu\nu}$ particle. However note that the helicity zero component of $w_{\mu\nu}$
does not correspond to a physical, propagating mode in the
massless limit, since it can be gauged away by the conformal transformations (\ref{conf}).
\vskip0.2cm
\noindent

\subsection{Higher tensors}

In this section we briefly discuss  some higher tensors, which are also of interest in the massless Weyl$^2$ theory.
In fact, the aim of this discussion is to construct conformal operators, which can be coupled to spin-four fields (see also the concluding section of this paper).
It is known in the literature that there are no totally symmetric, quadratic in the curvature and divergence-free  four-index tensors of dimension four \cite{BB} in a generic background. However, there are dimension four, divergence-free tensors which are totally symmetric  just in three-indices. It is also known that there exist a unique 
totally symmetric, traceless and divergence-free four-index tensor, on Ricci-flat spaces, which is the Bel-Robinson tensor \cite{BR}
\begin{eqnarray}
T_{\mu\nu\rho\sigma}&=&\frac{1}{4}\left(
\tensor{W}{^\lambda_\nu_\mu^\kappa}\tensor{W}{_\lambda_\sigma_\rho_\kappa}
+\frac{1}{2}\tensor{\epsilon}{^\lambda_\nu_\tau_\xi}
\tensor{\epsilon}{_\lambda_\sigma^\chi^\psi}\tensor{W}{^\tau^\xi_\mu^\kappa}
\tensor{W}{_\chi_\psi_\rho_\kappa}
 \right)\nonumber \\
 &=& \frac{1}{4}\left(  \tensor{W}{^\lambda_\nu_\mu^\kappa}\tensor{W}{_\lambda_\sigma_\rho_\kappa}+\tensor{W}{^\lambda_\sigma_\mu^\kappa}\tensor{W}{_\lambda_\nu_\rho_\kappa}-\frac{1}{2}
 g_{\nu\sigma} \tensor{W}{^\lambda^\tau_\mu^\kappa}\tensor{W}{_\lambda_\tau_\sigma_\kappa}
 \right).
 \label{t0}
\end{eqnarray}
This tensor is traceless $T^\mu_{~\mu\nu\rho}=0$ and satisfies
\begin{eqnarray}
 \nabla_\mu T^\mu_{~\nu\rho\sigma}= \frac{1}{2}\left({R_{\nu\kappa}^{~~\lambda}}_{\rho}
 \nabla_{[\lambda}R^\kappa_{~\sigma]}+
  {R_{\nu\kappa}^{~~\lambda}}_{\sigma}
 \nabla_{[\lambda}R^\kappa_{~\rho]}\right).
\end{eqnarray}
Hence in Einstein gravity, for Ricci-flat spaces, the Bel-Robinson tensor is 
divergence-free, $\nabla_\mu T^\mu_{~\nu\rho\sigma}=0$. Of course,
 Ricci-flat  manifolds are also Bach-flat but the contary is not true. For example, manifolds conformal to Einstein spaces have vanishing Bach tensor but non-vanishing Ricci.
In addition, the Bel-Robinson tensor transforms under Weyl rescalings as
\begin{eqnarray}
T^{\mu\nu}_{~~\rho\sigma}\to \widehat{T}^{\mu\nu}_{~~\rho\sigma}=
\Omega^{-4} T^{\mu\nu}_{~~\rho\sigma},
\end{eqnarray}
and therefore it has conformal dimension $\Delta_T=4$.\footnote{
The conformal dimension of a field in CFT  is the weight under the Weyl transformation of this field with half its indices up and half down. }
Let us recall here that in general, primary operators of spin $s$ and dimension $\Delta$  in $d$-dimensional unitary $CFT_d$, satisfy the unitarity bound \cite{Mack:1975je} 
 \begin{eqnarray}
\Delta\geq d-2+s.  \label{un}
\end{eqnarray}
In terms of the twist $\tau$ defined by 
\begin{eqnarray}
\tau=\Delta-s, 
\end{eqnarray}
the unitarity bound (\ref{un}) is written as
\begin{eqnarray}
\tau\geq d-2. 
\end{eqnarray}
Primary operators that saturate the bound satisfy the equation $\partial_{\mu_1} J^{\mu_1}_{~~\mu_2\cdots \mu_s}=0$ and therefore correspond to conserved operators. In particular, for $d=4$ we find that $\tau=2$ for a conserved primary  which is the case for the spin-two energy momentum tensor\footnote{In case
the energy momentum tensor is non-conserved and has dimension 
$\Delta>2+s$, the corresponding bulk spin-two field becomes massive \cite{Bachas}.}.  
However in our case the Bel-Robsinon operator has dimension $\Delta_T=4$ and spin $s=4$, and therefore it has twist $\tau_T=0$.
  It follows that the Bel-Robinson operator violates the unitarity bound, which is expected, since we know that the Weyl$^2$
contains ghosts and is therefore non-unitary.
Nevertheless as the ghost states  can be projected out by appropriate boundary conditions \cite{Maldacena},   we  may still look  for  a tensor  $J_{\mu\nu\rho\sigma}$ like the Bel-Robinson tensor that has the same symmetries with it,  it is divergence-free for Bach-flat spaces and has twist $\tau=2$. If such a tensor exist, 
  it 
  could couple to a  spin-4 field.
  In addition, the conformal dimension of $J_{\mu\nu\rho\sigma}$ should be 
\begin{eqnarray}
\Delta_J=6,
\end{eqnarray}
in order to describe a spin-4 conserved operator in the CFT. 
It is natural to expect that, similarly to the energy 
momentum tenor, $J_{\mu\nu\rho\sigma}$ is  quadratic in a tensor build out of  the curvature and/or its derivatives
of dimension $\Delta=3$. Such a tensor with $\Delta=3$ exists in conformal gravity  and it is  the Cotton tensor defined as 
\begin{eqnarray}
C_{\mu\nu\rho}=\nabla_\mu S_{\nu\rho}-\nabla_\nu S_{\mu\rho}, \label{cot}
\end{eqnarray}
where 
\begin{eqnarray}
S_{\mu\nu}=R_{\mu\nu}-\frac{1}{6} g_{\mu\nu} R,
\end{eqnarray}
is the Schouten tensor. 
In particular, by using the second Bianchi identity, the Cotton tensor can be written as the divergence of the Weyl tensor 
\begin{eqnarray}
C_{\mu\nu\rho}=\nabla_\kappa W^\kappa_{~\rho\mu\nu}, \label{cotW}
\end{eqnarray}
and therefore
\begin{eqnarray}
C_{[\mu\nu\rho]}=0. 
\end{eqnarray}

We can now define a conserved 4-tensor quadratic in the Cotton tensor as follows. 
From Eq.(\ref{cot}) we get the equation\footnote{To obtain Eq. (\ref{cc}), the relation $\nabla_\mu\nabla_\nu V^\rho-\nabla_\nu\nabla_\mu V^\rho
=R^{\rho}_{~\kappa\mu\nu}V^\kappa$ has been used.}
\begin{eqnarray}
\nabla_{[\lambda}C_{\rho\sigma] \kappa}= B_{\lambda\rho\sigma\kappa} \label{cc}
\end{eqnarray}
with 
\begin{eqnarray}
B_{\lambda\rho\sigma\kappa}=R^{\nu}_{~\kappa[\rho\lambda}S_{\sigma]\nu}=R^{\nu}_{~\kappa[ \rho\lambda }R_{\sigma]\nu}=W^{\nu}_{~\kappa[\rho\lambda}R_{\sigma]\nu},
\end{eqnarray}
whereas, using Eqs.(\ref{bach},\ref{cotW}) we find
\begin{eqnarray}
\nabla^\kappa C_{\kappa\mu\nu}=A_{\mu\nu},
\end{eqnarray}
with 
\begin{eqnarray}
A_{\mu\nu}=B_{\mu\nu}-\frac{1}{2}R^{\kappa\lambda}
W_{\mu\kappa\nu\lambda}.
\end{eqnarray}
Multiplying Eq.(\ref{cc}) by $C^{\lambda\rho\mu}$ and after some algebra
we arrive at 
\begin{eqnarray}
\nabla_\lambda J^{\lambda}_{~\alpha \mu\nu}=H_{\alpha \mu\nu},
\end{eqnarray}
where 
\begin{eqnarray}\label{jtensor}
J^{\lambda}_{~\alpha \mu\nu}=\frac{1}{4}\left(
C^{\lambda\gamma}_{~~\nu} C_{\alpha\gamma\mu}+C^{\lambda\gamma}_{~~\mu} C_{\alpha\gamma\nu}-\frac{1}{2} \delta^\lambda_\alpha \, C^{\rho\sigma}_{~~\nu} C_{\rho\sigma\mu}\right). 
\end{eqnarray}
and 
\begin{eqnarray}
H_{\alpha\mu\nu}=\frac{1}{8}\Big(C^{\gamma\lambda}_{~~\nu}B_{\alpha\lambda\gamma\mu}+C^{\gamma\lambda}_{~~\mu}B_{\alpha\lambda\gamma\nu}
+2A^\gamma_{~\mu}C_{\alpha\gamma\nu}+2
A^\gamma_{~\nu}C_{\alpha\gamma\mu}\Big).
\end{eqnarray}

Note that under a conformal transformation with $\delta g_{\mu\nu}=2\omega g_{\mu\nu}$, the Cotton tensor $C_{\mu\nu\rho}$ and the tensor $J_{\lambda\alpha \mu\nu}$ transform inhomogeneously under conformal   transformations. In particular the transformation of the Cotton tensor turns out to be  
\begin{eqnarray}
\delta C_{\mu\nu\rho}=2  \partial_\sigma \omega W^\sigma_{~\rho\mu\nu},
\end{eqnarray}
whereas, $J_{\lambda\alpha \mu\nu}$ transforms as 
\begin{eqnarray}
\delta J_{\lambda\alpha \mu\nu}=-2\omega J_{\lambda\alpha \mu\nu}
+\frac{1}{2}\partial_\tau \omega \left( \tensor{W}{^\tau^\gamma_\lambda_\nu}C_{\gamma\alpha\mu} +
\tensor{W}{^\tau_\gamma_\alpha_\nu}C^{\gamma}_{~\lambda\mu}-\frac{1}{2}\tensor{W}{^\tau^\rho^\sigma_\nu}C_{\rho\sigma\mu}+\mu\leftrightarrow \nu
\right).
\end{eqnarray}
This means that both  $C_{\mu\nu\rho}$ and $J_{\lambda\alpha \mu\nu}$, although not Weyl invariant,  are Weyl covariant \cite{Boulanger:2004eh} as their transformation under Weyl rescalings does not involve higher than first derivatives of the conformal factor.
 Therefore in order to obtain a 
conformal tensor one should use a Weyl-covariant derivative for the construction of an ``improved" Cotton tensor, along the lines 
of    \cite{Boulanger:2004eh}. 

\noindent
 It is easy to prove that in the linearized theory, the $J^{\lambda\mu\nu\rho}$  is  divergence-free and has the correct conformal  dimension $\Delta_J=6$
\begin{eqnarray}
&& \partial_\mu \tensor{J}{^\mu^\nu_\rho_\sigma}=0, ~~~\nonumber \\
 &&\widehat{J}^{\mu\nu}_{~~\rho\sigma}=\Omega^{-6}
 \tensor{J}{^\mu^\nu_\rho_\sigma}. \label{jj}
 \end{eqnarray} 
If (\ref{jj})  can be extended  at the full non-linear level is not known to us.

\section{${\cal N}=1$ Super-Weyl theory}

\subsection{Massive theory}

Now we want to present the superfield versions of the bosonic Weyl$^2$-action, first using ${\cal N}=1$ supersymmetry language. 
In conformal supergravity, one 
first introduces the super-Weyl tensor ${\cal W}_{\alpha\beta\gamma}$, which is a chiral superfield, where $\alpha,\beta,\gamma=1,2$ are standard $SL(2,\mathbf{C})$ spinor indices.
${\cal W}_{\alpha\beta\gamma}$ has spin $({3\over 2},0)$, i.e. its highest component is a fermionic spin-3/2 field, but ${\cal W}_{\alpha\beta\gamma}$ also contains
the spin-two field $g_{\mu\nu}$ with five degrees of freedom. 
The Weyl superfield ${\cal W}_{\alpha \beta \gamma}$ satisfies
\be
{\cal W}_{\alpha \beta \gamma} = {\cal W}_{(\alpha \beta \gamma)} \  , \  
({\cal W}_{\alpha \beta \gamma})^* = \overline {\cal W}_{\dot \alpha \dot \beta \dot \gamma} \  , \   
\overline{ {\cal D}}_{\dot{\delta}} 
{\cal W}_{\alpha \beta \gamma}= 0  \ ,  \label{WW}
\ee
and it has the
 following pure bosonic contributions 
\be
{\cal D}_{\gamma} {\cal W}_{\delta \epsilon \alpha} |=  
\frac16 \left( \frac{i}{4}  \epsilon_{\gamma \delta} {\cal D}_{\epsilon \dot \epsilon} A_{\alpha}^{\dot \epsilon} 
-\frac{1}{4} 
\epsilon^\beta_{~\epsilon} \epsilon^\rho_{~ \gamma} 
W_{\delta\beta\alpha\rho}
%
\right) + (\delta \, \epsilon \, \alpha) \, ,  
\ee
where $W_{\delta\beta\alpha\rho}$ and $A_{\alpha}^{\dot \epsilon}$ are the spinorial equivalent or the Weyl conformal tensor and the vector auxiliary of the ${\cal N}=1$ supergravity, 
 and $(\delta \, \epsilon \, \alpha)$ 
denotes five terms obtained by symmetrization with respect to the fermionic 
indices $\delta$, $\epsilon$ and $\alpha$ \cite{FFKL}.

The supersymmetric action of massive Weyl gravity is then written as 
\be
\label{fw1}
{\cal L}_{\Phi,{\cal W}} =  \int d^2 \Theta \, 2{\cal E} {\cal R} +    4 \int d^2 \Theta \, 2{\cal E}
\, \tau \, {\cal W}^2
+c.c., 
\ee
where $\tau$ is  the complex coupling
\begin{eqnarray}
\tau=\frac{1}{g^2}+i\alpha.
\end{eqnarray}
We then find that the bosonic sector of the action is
\be
\label{fw2}
e^{-1} {\cal L}_{{\cal W}} =  \frac{1}{2}R+\frac{1}{3} A_\mu A^\mu-\frac{1}{3} u\overline u+
\frac{\alpha}{2} 
\left( \frac12 R^2_{HP}
-\frac{2}{3} F_{\mu\nu} F_{\rho\sigma} \epsilon^{\mu\nu\rho\sigma}
\right) +
\frac{1}{2g^2}  \left( W^{\mu\nu\rho\sigma}W_{\mu\nu\rho\sigma}
- \frac43 F^{\mu\nu} F_{\mu\nu}
\right) 
\, , 
\ee
where $F_{\mu\nu}=\partial_\mu A_\nu-\partial_\nu A_\mu$ is the field strength of the supergravity vector $A_\mu$, $u$ is the auxiliary scalar and $R^2_{HP} = W_{\mu\nu \kappa\lambda} \epsilon^{\kappa\lambda \rho\sigma} W_{\rho\sigma}^{\ \ \mu\nu}$ is the topological Hirzebruch--Pontryagin  term. 
Note that in Weyl supergravity,
the vector $A_\mu$ is dynamical as it has  a  kinetic term of the form $ F^{\mu\nu} F_{\mu\nu }$.

\vskip0.6cm
\noindent
Now we are ready to determine 
 the spectrum of the massive ${\cal N}=1$ Super-Weyl theory. It has the following form:
\vskip0.5cm
\noindent
(i) A  standard massless spin-two graviton multiplet with the following $(h,q_R)$ helicity $h$ components and $U(1)_R$ $q_R$  charges and with $n_B+n_F=4$ degrees of freedom:
\begin{eqnarray}
g_{{\cal N}=1}:~   (+2,0) +  (+{3\over2},+{1\over 2} )    \, ,
\end{eqnarray}
and its CPT conjugate 
\begin{eqnarray}
 (-{3\over2},-{1\over 2} )+(-2,0)    \, .
\end{eqnarray}

\vskip0.2cm
\noindent
(ii)  In the non-standard sector there is a massive spin-two supermultiplet,\footnote{General massive multiplets in extended supersymmetry
were discussed in \cite{Ferrara:1980ra}.}  which is the socalled massive ${\cal N}=1$ super-Weyl multiplet 
  \cite{Ferrara:1977mv}
with $n_B+n_F=16$ degrees of freedom:
\begin{eqnarray}
w_{{\cal N}=1}:~    {\rm Spin}(2) + {\underline{2}}\times {\rm Spin} ( 3/2 )  + 
 {\rm Spin}(1)       \, .
\end{eqnarray}
Note that the massive states in ${\cal N}=1$ supergravity built representations of the group $USp(2)$.
\vskip0.2cm
\noindent
Hence in summary, the massive super-(Weyl)$^2$ gravity theory contains $n_B+n_F=20$ degrees of freedom.

\subsection{Massless theory}

The Weyl equation of motion eq.(\ref{weq}) now reads at linearized level 
\cite{FZ}:
\begin{equation}\label{superw}
D^\alpha\partial^\beta_{\dot \beta}{\cal W}_{\alpha\beta\gamma}=0\, .
\end{equation}
This equation is equivalent to the linearized Weyl equation of motion of the bosonic Weyl tensor, as given in eq.(\ref{linWeyl}).
Then the spectrum of the massless ${\cal N}=1$ Super-Weyl theory has the following form:
\vskip0.5cm
\noindent
(i) A  standard massless spin-two supergravity multiplet with $n_B+n_F=4$ degrees of freedom and the following $(h,q_R)$ helicity $h$ components and $U(1)_R$ $q_R$  charges:
\begin{eqnarray}
g_{{\cal N}=1}:~    (+2,0) +   (+{3\over2},+{1\over 2})     \, ,
\end{eqnarray}
together with its CPT conjugate multiplet
\begin{eqnarray}
  (-{3\over2},-{1\over 2} )+(-2,0)   . \label{cpt1}
\end{eqnarray}

Note that the $U(1)_R$  charges and the helicities of the fermions are correlated, i.e. the states with positive helicity have also positive $U(1)_R$  charge,
and due to CPT the opposite is true for the states with negative helicities.

\vskip0.2cm
\noindent
(ii)  In the non-standard sector,   we decompose the massive states into their massless helicity components.\footnote{Note  when decomposing massive into massless multiplets, the latter always come with their CPT conjugate multiplet
because the massive ones are CPT invariant.}
Furthermore we have to decompose the $USp(2)$ representations into the $U(1)_R$ charges of the massless states:
\begin{equation}
USp(2)\,\rightarrow \, U(1)_R:\quad{\underline 2}={1\over 2}\oplus-{1\over 2}\, ,
\end{equation}
First, we get   from the massive Weyl multiplet  $w_{{\cal N}=1}$ a massless ghost-like spin-two supermultiplet: 
\begin{eqnarray}
{\rm spin-two}:~    (+2,0) +   ( +{3\over2},+{1\over 2} )  \, ,
\end{eqnarray}
and its CPT conjugate as in   (\ref{cpt1}).
Second we get from $w_{{\cal N}=1}$ a massless, physical spin-3/2 supermultiplet:
\begin{eqnarray}
{\rm spin-3/2}:~    (+{3\over2},-{1\over 2}) +  (+1,0)   \, , \label{32v}
\end{eqnarray}
together with its CPT conjugate multiplet 
\begin{eqnarray}
(-1,0) +(-{3\over2},+{1\over 2})   \, . \label{32v1}
\end{eqnarray}
The spin-3/2 fields together build 
a so-called 
tripole ghost, which effectively acts as a physical spin-3/2 multiplet and a dipole ghost spin-2 multiplet \cite{FZ}.

And thirdly,  $w_{{\cal N}=1}$ provides a physical, massless spin-one vector multiplet:
\begin{eqnarray}
{\rm spin-one}:~    (+1,0) +  ( +{1\over2},+{1\over 2})     \, , \label{v11}
\end{eqnarray}
and its CPT conjugate 
\begin{eqnarray}
  ( -{1\over2},-{1\over 2})  +(-1,0)    \, . \label{v12}
\end{eqnarray}
Note that the massive Weyl multiplet  $w_{{\cal N}=1}$ contains in addition a chiral spin-1/2 multiplet:
  \begin{equation} 
  (+{1\over 2},-{1\over 2}) +  ( 0,0)   \, , 
  \end{equation}
  and its CPT conjugate
  \begin{equation} 
    ( 0,0 ) + (-{1\over 2},+{1\over 2})   \, . 
  \end{equation}
However this multiplet is unphysical since it can be gauge away by the superconformal transformations together with the local $U(1)_R$ transformations.
Specifically, one of the two scalars in this chiral multiplet is the Weyl mode, i.e. the helicity zero component of the massive spin-two field $w_{\mu\nu}$ in $w_{{\cal N}=1}$
and the other scalar is the helicity 
zero component of the massive vector inside $w_{{\cal N}=1}$, which is gauged away by the  $U(1)_R$ transformations.

\vskip0.2cm
\noindent
Hence in summary, the massless super-(Weyl)$^2$ gravity theory contains $n_B+n_F=16$ physical, propagating degrees of freedom \cite{LvN}.
Regarding the pure super-Weyl lagrangian in Eq.(\ref{fw2}), when the standard sugra term is omitted, the graviton $\pm1$ states belong to the ${\cal N}=1$ vector multiplet (Eqs.(\ref{v11}) and (\ref{v12})), while the $A_\mu$  field is the gauge field of the $U(1)$ R-symmetry\ and belongs to the spin-3/2 multiplet (Eqs.(\ref{32v}) and (\ref{32v1})).

We note that, in accordance to \cite{LvN}, the gravitino action contains 
(${\cal N}=1$) eight helicity states, three (tripole ghost) $(+\frac{3}{2},+\frac{1}{2}),(+\frac{3}{2},+\frac{1}{2}),(+\frac{3}{2},-\frac{1}{2})$ and a spin-1/2 state $(+\frac{1}{2},+\frac{1}{2})$ (+  CPT conjugates). 
This phenomenon will be present for all ${\cal N}$, so the spectrum always contains the states
$
(+\frac{3}{2},\mathcal{N})~(+\frac{3}{2},\mathcal{N}),(+\frac{3}{2},\overline{\mathcal{N}}), (+\frac{1}{2},\mathcal{N})+\mbox{CPT conjugates},
$
(the last comes from the gravitino multiplet for ${\cal N}=4)$.
For ${\cal N}=4$ there are additional spin-1/2 states in the $4,\overline{4}$ and $\overline{20}$ of $SU(4)$. The three states $(+\frac{1}{2},\overline{4}),(+\frac{1}{2},\overline{4}),(+\frac{1}{2},4)$ (the first two from the spin-2 multiplets and the last from the spin-3/2 multiplet) and their CPT conjugates, are described by a spin-1/2 cubic kinetic term, and finally the $(+\frac{1}{2},\overline{20}) (+\mbox{CPT conjugate})$ is described by a standard (first derivative) kinetic term \cite{Bergshoeff:1980is}.

 \subsection{Higher tensors}

Again, we like to consider  some higher tensor in the massless ${\cal N}=1$ super-Weyl$^2$ theory.
We will restrict ourselves to construct the relevant superfields in the linearized approximation of the super-Weyl$^2$ theory.
First,
 the linearized super-Bel Robinson tensor has the following form:
\begin{equation}
{\cal T}_{\alpha\beta\gamma,\dot\alpha\dot\beta\dot\gamma}={\cal W}_{\alpha\beta\gamma}
\overline {\cal W}_{\dot\alpha\dot\beta\dot\gamma}\, .
\end{equation}
It is a tensor of spin $({3\over 2},{3\over 2})$ and contains a bosonic spin-four field.  The bosonic part of the super-Bel Robinson tensor is 
\begin{eqnarray}
T_{\alpha\beta\gamma\delta,\dot\alpha\dot\beta\dot\gamma\dot{\delta}}=D_\delta \overline D_{\dot{\delta}}{\cal T}_{\alpha\beta\gamma,\dot\alpha\dot\beta\dot\gamma}|=W_{\alpha\beta\gamma\delta}\overline  W_{\dot\alpha\dot\beta\dot\gamma\dot{\delta}}.
\end{eqnarray}
Recalling that $W_{\alpha\beta\gamma\delta}$ and $ W_{\dot\alpha\dot\beta\dot\gamma\dot{\delta}}$ correspond to the anti-self dual ${}^-W^\lambda_{~~\nu\mu\kappa}$ and the self dual ${}^+W^\lambda_{~~\nu\mu\kappa}$ parts of the Weyl tensor, we find the components of the Bel-Robinson tensor with Lorentz indices $T_{\mu\nu\rho\sigma}$ to be
\begin{eqnarray}
T_{\mu\nu\rho\sigma}={}^-W^\lambda_{~~\nu\mu\kappa}~{}^+W_{\lambda\sigma\rho}^{~~~\kappa}, \label{tt}
\end{eqnarray}
where 
\begin{eqnarray}
{}^\pm W_{\mu\nu\rho\sigma}=\frac{1}{2}\Big( W_{\mu\nu\rho\sigma}\mp
  i \,{}^*W_{\mu\nu\rho\sigma}
\Big),
\end{eqnarray}
with 
\begin{eqnarray}
{}^*W_{\mu\nu\rho\sigma}=\frac{1}{2}\epsilon_{\mu\nu}^{~~\kappa\lambda}
W_{\kappa\lambda\rho\sigma}. \label{dual}
\end{eqnarray}
Therefore, we may express (\ref{tt}) as
\begin{eqnarray}
T_{\mu\nu\rho\sigma}=\frac{1}{4}\Big({W^{\lambda}}_{\nu\mu}^{~~\kappa}\,
W_{\lambda\sigma\rho\kappa}+{}^*{W^{\lambda}}_{\nu\mu}^{~~\kappa}{}^*W_{\lambda\sigma\rho\kappa}\Big), \label{ttt}
\end{eqnarray}
which coincides with the definition (\ref{t0}).
Notice that due to eq.(\ref{WW}) the super-Bel Robinson superfield  
${\cal T}_{\alpha\beta\gamma,\dot\alpha\dot\beta\dot\gamma}$ satisfies
\begin{eqnarray}
 D^\alpha {\cal T}_{\alpha\beta\gamma,\dot\alpha\dot\beta\dot\gamma}=
 \overline D^{\dot \alpha}{\cal T}_{\alpha\beta\gamma,\dot\alpha\dot\beta\dot\gamma}=0, 
 \end{eqnarray}
 and therefore, it is  conserved in Einstein supergravity.
 \begin{eqnarray}
  \partial^{\alpha\dot\alpha}{\cal T}_{\alpha\beta\gamma,\dot\alpha\dot\beta\dot\gamma}=0.
  \end{eqnarray} 
   However in Weyl supergravity, 
this is not the case (at least from Bach-flat but not Ricci-flat backgrounds) and therefore 
$T_{\alpha\beta\gamma,\dot\alpha\dot\beta\dot\gamma}$ is not the correct object we are looking for, as in addition it has dimension four.

As in the bosonic case, we are considering the  super-Cotton tensor, which is given as 
\begin{equation}
{\cal C}_{\dot\alpha\beta\gamma}=\partial^\alpha_{\dot\alpha}
{\cal W}_{\alpha\beta\gamma}\, .
\end{equation}
Note that ${\cal C}_{\dot\alpha\beta\gamma}$ is symmetric in the  last two indices $\beta$ and $\gamma$ and it has spin $(1,{1\over 2})$. 
Using the super-Cotton tensor the linear
super-Weyl equation of motion (\ref{superw}) can be rewritten as
\begin{equation}\label{superwa}
D^\beta {\cal C}_{\dot\alpha\beta\gamma}=0\, .
\end{equation}

Finally we construct the super-version of the bosonic tensor $J^{\lambda}_{~\alpha \mu\nu}$ in eq.(\ref{jtensor}). It has the following form
\begin{equation}\label{superj}
{\cal J}_{\dot\alpha\beta\gamma,\alpha\dot\beta\dot\gamma}={\cal C}_{\dot\alpha\beta\gamma}\overline{\cal  C}_{\alpha\dot\beta\dot\gamma}\, .
\end{equation}
The tensor $J_{\dot\alpha\beta\gamma,\alpha\dot\beta\dot\gamma}$ has spin $({3\over 2}+{1\over 2},{3\over 2}+{1\over 2})$. When writing this
field in components, it contains a spin-four field, namely precisely $J^{\lambda\mu\nu\rho}$, which was introduced before. Indeed,
we have that 
\begin{eqnarray}
J_{\dot\alpha\beta\gamma\delta,\alpha\dot\beta\dot\gamma\dot\delta}
=D_\delta \overline D_{\dot\delta} {\cal J}_{\dot\alpha\beta\gamma,\alpha\dot\beta\dot\gamma}|=C_{\dot\alpha\beta\gamma\delta}\overline C_{\alpha\dot\beta\dot\gamma\dot\delta}=\partial_{\dot \alpha}^\alpha W_{\alpha\beta\gamma\delta}\partial_{\alpha}^{\dot\alpha} W_{\dot\alpha\dot\beta\dot\gamma\dot\delta}.
\end{eqnarray}
Proceeding as above, we find that the corresponding Lorentz index tensor 
$J_{\nu\sigma\mu\rho}$ is given by 
\begin{eqnarray}
J_{\nu\sigma\mu\rho}=\partial^\lambda {}^-W^\kappa_{~~\nu\mu\lambda}~
\partial^\tau {}^+W_{\kappa\sigma\rho\tau}
=\frac{1}{4}\Big(\partial^\lambda W^\kappa_{~~\nu\mu\lambda}~
\partial^\tau W_{\kappa\sigma\rho\tau}+\partial^\lambda {{}^*W}^\kappa_{~~\nu\mu\lambda}~
\partial^\tau {{}^*W}_{\kappa\sigma\rho\tau}\Big), 
\end{eqnarray}
which can be written in terms of the Cotton tensor as 
\begin{eqnarray}
J_{\nu\sigma\mu\rho}&=&   \frac{1}{4}
\Big(C^\kappa_{~\nu\mu} C_{\kappa\sigma \rho}+\frac{1}{4}
\tensor{\epsilon}{^\kappa_\nu^\lambda^\xi}
\tensor{\epsilon}{_\kappa_\sigma^\chi^\psi} C_{\lambda\xi\mu}
\tensor{C}{_\chi_\psi_\rho}\Big)\nonumber \\
&=& \frac{1}{4}\Big(C^{\kappa}_{~\nu\mu} C_{\kappa\sigma\rho}+
C^{\kappa}_{~\nu\rho} C_{\kappa\sigma\mu}-\frac{1}{2} g_{\nu\sigma}
 \, C^{\kappa\lambda}_{~~\mu} C_{\kappa\lambda\rho} \Big),\label{jjtensor}
\end{eqnarray} 
i.e., coincides with 
Eq.(\ref{jtensor})

However note that
$J_{\dot\alpha\beta\gamma,\alpha\dot\beta\dot\gamma}$ is not irreducible in terms of Lorentz spins, but it also contains fields of lower spins:
\begin{equation}
J_{\dot\alpha\beta\gamma,\alpha\dot\beta\dot\gamma}:\quad 3+3(2)+4(1)+2(0)\, .
\end{equation}
Using the super-Weyl equation of motion one can show that $J_{\dot\alpha\beta\gamma,\alpha\dot\beta\dot\gamma}$ is indeed covariantly conserved:
\begin{equation}\label{superjcons}
D^\beta J_{\dot\alpha\beta\gamma,\alpha\dot\beta\dot\gamma}=D^{\dot\beta} J_{\dot\alpha\beta\gamma,\alpha\dot\beta\dot\gamma}=0\, .
\end{equation}
\noindent

Equation (\ref{superjcons}) can be rewritten as
\begin{equation}
D^\beta J_{\dot\alpha\beta\gamma,\alpha\dot\beta\dot\gamma}=D_{\dot\alpha} S_{\dot\beta\gamma,\alpha\dot\gamma}=0\, .
\end{equation}
$S_{\dot\beta\gamma,\alpha\dot\gamma}$ is a superfield with highest spin-2 component (and containing a bosonic spin-3 field), namely its spin content is as follows:
\begin{equation}
S_{\dot\beta\gamma,\alpha\dot\gamma}:\quad           ({1\over 2},{1\over 2})     \otimes   ({1\over 2},{1\over 2}) =  2+3(1)+2(0)\, .
\end{equation}
Therefore 
this conservation equation implies that these  components
in $J_{\dot\alpha\beta\gamma,\alpha\dot\beta\dot\gamma}$ get eliminated and the physical spin-components of $J_{\dot\alpha\beta\gamma,\alpha\dot\beta\dot\gamma}$ are the
following:
\begin{equation}
J_{\dot\alpha\beta\gamma,\alpha\dot\beta\dot\gamma}:\quad 3+2(2)+1\, .
\end{equation}

\section{${\cal N}=2$ Super-Weyl theory}

\subsection{Massive theory}

The spectrum of the massive ${\cal N}=2$ Super-Weyl theory has the following form:
\vskip0.5cm
\noindent
(i) A  standard massless spin-two super graviton multiplet with $n_B+n_F=8$ degrees of freedom. 
It contains states with the following helicities and  their associated $U(2)$ quantum numbers:
\begin{eqnarray}\label{s2a}
 (2_1,\underline 1_0) + ( {3\over 2},\underline 2_1 ) +(1,\underline 1_2)
  \, .
\end{eqnarray}
In addition there are the following CPT conjugate states
\begin{eqnarray}\label{s2b}
 (-1,\underline 1_{-2}) + ( -{3\over 2},\underline 2_{-1} ) +(-2,\underline 1_0)
  \, .
\end{eqnarray}

\vskip0.2cm
\noindent
(ii)  In the non-standard sector we have a massive spin-two supermultiplet, which is the ${\cal N}=2$ massive super-Weyl multiplet \cite{deWit:1979dzm} with $n_B+n_F=48$ degrees of freedom:
\begin{eqnarray}
w_{{\cal N}=2}:~    {\rm Spin}(2) + {\underline{4}}\times {\rm Spin} ( 3/2 )  + ({\underline{5}}+{\underline{1}})
\times {\rm Spin}(1) + {\underline{4}}\times {\rm Spin} (1/2) +  {\rm Spin} (0)        \, .
\end{eqnarray}
Note that the massive states in ${\cal N}=2$ supergravity built representations of the group $USp(4)$.
\vskip0.2cm
\noindent
Hence in summary, the ${\cal N}=2$ massive super-(Weyl)$^2$ gravity theory contains $n_B+n_F=56$ degrees of freedom.

\subsection{Massless theory}

Specifically the spectrum of the massless ${\cal N}=2$ Super-Weyl theory has the following form:
\vskip0.5cm
\noindent
(i) A  standard massless spin-two super graviton multiplet with $n_B+n_F=8$ degrees of freedom. Its helicity and $U(2)$ quantum numbers are as in eqs.(\ref{s2a}) and (\ref{s2b}).

\vskip0.2cm
\noindent
(ii)  In the non-standard sector,   we get first  from the massive Weyl multiplet  $w_{{\cal N}=2}$ a massless ghost-like spin-two supermultiplet with $n_B+n_F=8$.
It contains  states with  helicity and  $U(2)$ quantum numbers, again as given in eqs.(\ref{s2a}) and (\ref{s2b}).

Second we get from $w_{{\cal N}=2}$ two massless spin-3/2 supermultiplets with in total $n_B+n_F=16$ degrees of freedom:
\begin{eqnarray}
{{\underline{{ 2}}}}_{-1}\times\lbrack   ({3\over 2},\underline 1_0) +  ( 1,\underline2_1 )+({1\over2},\underline 1_2)
\rbrack     \, .
\end{eqnarray}
As before, there are the following additional CPT conjugate states
\begin{eqnarray}
{{\underline{{ 2}}}}_{1}\times\lbrack  
(-{1\over2},\underline 1_{-2})
+  ( -1,\underline2_{-1} )
 +(-{3\over 2},\underline 1_0) 
\rbrack     \, .
\end{eqnarray}
Altogether they contain the four gauge bosons of the local $U(2)_R$ gauge symmetry in the $\underline 1_0+\underline 3_0$ representations.

 The massive Weyl multiplet  $w_{{\cal N}=2}$ contains in addition two ${\cal N}=2$ vector multiplets.
 The first vector multiplet has the form:
 \begin{equation}
  (1,\underline1_0) +  ( {1\over 2},\underline2_1 )+
( 0,\underline 1_2 )\, .
\end{equation}
Its CPT conjugate multiplet is given as
\begin{equation}
 ( 0,\underline 1_{-2} )
 +  ( -{1\over 2},\underline2_{-1} )
 +
    (-1,\underline1_0) \, .
\end{equation}
This vector multiplet is physical and contains propagating degrees of freedom.
The neutral massless vector in this equation is the original $\pm 1$ helicity partner
of the graviton.

The second vector multiplet and its CPT conjugate possess different $U(1)$ charges, namely
 \begin{equation}
  (1,\underline1_{-2}) +  ( {1\over 2},\underline2_{-1} )+
( 0,\underline 1_0 )
 \, ,
\end{equation}
and 
\begin{equation}
 ( 0,\underline 1_{0} )
 +  ( -{1\over 2},\underline2_{1} )
 +
    (-1,\underline1_2)\ \, .
\end{equation}
It contains a neutral scalar, namely  the Weyl mode,
which is gauged away by the conformal transformations.
Hence this vector-multiplet gets removed, it is unphysical and does not propagate.

Finally the non-standard sector also contains one CPT self-conjugate, massless ${\cal N}=2$ hyper multiplet:
  \begin{equation}
    ( {1\over 2 },\underline 2_{-1})+
 ( 0 ,\underline 2_{-1}\otimes 2_{1})+ ( -{1\over 2 },\underline 2_{1})
 \, .
\end{equation}
This hyper multiplet is unphysical,  namely the four scalars in the  $\underline 1_0+\underline 3_0$ representations
are the helicity 
zero components of the massive vectors inside $w_{{\cal N}=2}$, which are gauged away by the local $U(2)_R$ transformations.


\vskip0.2cm
\noindent
Hence in summary, the massless ${\cal N}=2$ super-(Weyl)$^2$ gravity theory contains $n_B+n_F=40$ physical, propagating degrees of freedom.

\section{${\cal N}=3$ Super-Weyl theory}

\subsection{Massive theory}

The spectrum of the massive ${\cal N}=3$ Super-Weyl theory has the following form:
\vskip0.5cm
\noindent
(i) 
A  standard massless spin-two super graviton multiplet $g_{{\cal N}=3}$ with $n_B+n_F=16$ degrees of freedom and the following helicity and $U(3)$ quantum numbers:
\begin{eqnarray}\label{gn3a}
    ( 2,\underline 1_0) + ( {3\over 2},\underline 3_1 )+ ( 1,\bar{\underline 3}_{2} )+ 
( {1\over2},\underline 1_3 )\, .
  \end{eqnarray}
In addition one obtains the CPT conjugate states:
\begin{eqnarray}\label{gn3b}
   ( - {1\over2},\underline 1_{-3} )+  ( -1,{\underline 3}_{-2} ) +  ( -{3\over 2},\bar{\underline 3} _{-1})+(- 2,\underline 1_0)
  \, .
\end{eqnarray}

\vskip0.2cm
\noindent
(ii)  In the non-standard sector we have a massive spin-two supermultiplet, which is the ${\cal N}=3$ massive super-Weyl multiplet with $n_B+n_F=128$ degrees of freedom  \cite{vanMuiden:2017qsh}:
    \begin{eqnarray}     
w_{{\cal N}=3}:   
 {\rm Spin}(2) + {\underline{6}}\times {\rm Spin} ( 3/2 )  + ({\underline{14}}+
{\underline{1}})\times {\rm Spin}(1) + ({\underline{14'}}+{\underline{6}})\times {\rm Spin} (1/2) 
+ {\underline{14}}\times {\rm Spin} (0)      \,.
\end{eqnarray}
Note that the massive states in ${\cal N}=3$ supergravity built representations of the group $USp(6)$.
\vskip0.2cm
\noindent
Hence in summary, the ${\cal N}=3$ massive super-(Weyl)$^2$ gravity theory contains $n_B+n_F=144$ degrees of freedom.

\subsection{Massless theory}

The  spectrum of the massless ${\cal N}=3$ Super-Weyl theory has the following form:
\vskip0.5cm
\noindent
(i) A  standard massless spin-two super graviton multiplet with $n_B+n_F=16$ degrees of freedom, as given in eqs.(\ref{gn3a}) and (\ref{gn3b}).

\vskip0.2cm
\noindent
(ii)  In the non-standard sector, 
in order to obtain the massless states, we have to decompose the
 massive representations into massless representations. This is done via the branching rules of the massive $USp(6)$ R-symmetry group
 into the R-symmetry group $U(3)$ of the massless states. The specific decomposition of $USp(6)\rightarrow SU(3)\times U(1)$ for the relevant representations is as follows:
\begin{eqnarray}
 {\underline{6}}&=&      { {\underline{{ 3}}}}_1\oplus{\bar {\underline{{ 3}}}} _{-1}    \, ,  \nonumber\\
{\underline{14}}&=&     { {\underline{{ 3}}}}_{-2}\oplus{\bar {\underline{{ 3}}}}_{2}\oplus{ {\underline{{ 8}}}}_0\, ,     \nonumber\\
{\underline{14'}}&=& { {\underline{{ 1}}}}_{3}\oplus{ {\underline{{ 1}}}}_{-3}\oplus { {\underline{{ 6}}}}_{-1}\oplus{\bar {\underline{{ 6}}}}_1\, . 
\end{eqnarray}

\vskip0.4cm\noindent
Then we get first  from the massive Weyl multiplet  $w_{{\cal N}=3}$ a         spin-two supermultiplet, again with the states as given eqs.(\ref{gn3a}) and (\ref{gn3b}). 
They constitute  a massless ghost-like spin-two supermultiplet with $n_B+n_F=16$.

Second we get from $w_{{\cal N}=3}$  a helicity +3/2 supermultiplet with the  $U(3)$ charge assignements
\begin{eqnarray}
{\bar{\underline 3}_{-1}}\otimes\Bigl\lbrack
( {3\over 2},{\underline 1}_{0} ) +( 1,\underline 3 _1)
 +( {1\over2},{\underline 3}_2 )
 +( 0,\underline 1_{3} )\Bigr\rbrack
 \, ,
\end{eqnarray}
plus the CPT conjugate multiplet of the form
\begin{eqnarray}
{{\underline 3}_{1}}\otimes\Bigl\lbrack
 ( 0,\underline 1_{-3} )+
 ( -{1\over2},\bar{\underline 3}_{-2} )+
 ( -1,\bar{\underline 3} _{-1})+
( -{3\over 2},{\underline 1}_{0} ) 
  \Bigr\rbrack
 \, .
\end{eqnarray}
These two multiplets together build three massless, physical spin-3/2 supermultiplets with in total $n_B+n_F=48$.
They contain the nine gauge bosons, which transform as $\underline 1_0+\underline 8_0$,  of the local $U(3)_R$ gauge symmetry.

Finally the massive Weyl multiplet  $w_{{\cal N}=3}$ contains in addition the following helicity states:
\begin{equation}\label{v1}
({\underline 1}_0+{\underline 3}_{-2})\otimes\Bigl\lbrack
  ( 1,{\underline 1}_0 )
  +( {1\over2},{\underline 3}_1 )
  +( 0,\bar{\underline 3}_2)
 + (- {1\over2},\underline 1_{3} )\Bigr\rbrack
  \, .
\end{equation} 
  There are also  the CPT conjugate states of the form: 
  \begin{equation}\label{v2}
 ({\underline 1}_0+\bar{\underline 3}_{2})\otimes\Bigl\lbrack
 ({1\over2},\underline 1_{-3} )
 +( 0,{\underline 3}_{-2})
  +( -{1\over2},{\bar{\underline 3}_{-1}} )
 + ( -1,{\underline 1}_0 )
  \Bigr\rbrack
 \, .
\end{equation}

Altogether the states in eqs.(\ref{v1}) and (\ref{v2}) built four full, massless ${\cal N}=3$ vectormultiplets.
The 24 scalars in these four vector multiplets transform under $U(3)$ as ${ {\underline{{ 1}}}}_0+{{\underline{{ 1}}}}_0+{ {\underline{{ 3}}}}_{-2}+{\bar {\underline{{ 3}}}}_2+{{\underline{{ 8}}}}_0+{{\underline{{ 8}}}}_0$.
One singlet scalar is gauged away by the conformal transformations. Furthermore nine scalars in  ${ {\underline{{ 1}}}}_0+{{\underline{{ 8}}}}_0$ representations 
are the helicity 
zero component of the massive vectors inside $w_{{\cal N}=3}$, which are gauged away by the local $U(3)_R$ transformations.
 It follows that three of the four vector multiplets get removed. They are unphysical and correspond to superconformal and $U(3)$ gauge degrees of freedom.
Therefore one gets just one physical, propagating, massless ${\cal N}=3$ vector multiplet with the following fields:
\begin{equation}\label{v111}
  ( 1,{\underline 1}_0 )
  +( {1\over2},{\underline 3}_1 )
  +( 0,\bar{\underline 3}_2)
 + (- {1\over2},\underline 1_{3} )
  \, ,
\end{equation} 
together with its CPT conjugate state.

\vskip0.2cm
\noindent
Hence in summary, the massless ${\cal N}=3$ super-(Weyl)$^2$ gravity theory contains $n_B+n_F=96$ physical, propagating degrees of freedom.

\section{${\cal N}=4$ Super-Weyl theory}

\subsection{Massive theory}

The spectrum of the massive ${\cal N}=4$ Super-Weyl theory has the following form:
\vskip0.5cm
\noindent
(i) A  standard massless spin-two super graviton multiplet $g_{{\cal N}=4}$ with $n_B+n_F=32$ degrees of freedom
and with the following helicities  and $SU(4)$ representations:
\begin{eqnarray}\label{gn4a}
   (+2,\underline1) + ( +{3\over 2} ,\underline 4) +(1,\underline 6) + ( +{1\over2},\underline{\overline4} )   + ( 0,\underline1)     \, ,
\end{eqnarray}
together with its CPT conjugate 
\begin{eqnarray} \label{gn4b}
  ( 0,\underline1)   + ( -{1\over2},\underline4 )  +(-1,\underline 6) 
   + ( -{3\over 2} ,\underline {\overline 4}) +(-2,\underline1) .
\end{eqnarray}

\vskip0.2cm
\noindent
(ii)  In the non-standard sector we have 
the spin-two massive Weyl multiplet of ${\cal N}=4$, which is irreducible with $n_B+n_F= 2^8=256$
 with states in $USp(8)$ representations       \cite{Bergshoeff:1980is}:
 \begin{eqnarray}
w_{{\cal N}=4}:~&   
{\rm Spin}(2) + {\underline{8}}\times {\rm Spin} ( 3/2 )  + {\underline{27}}
\times {\rm Spin}(1) + {\underline{48}}\times {\rm Spin} (1/2) + {\underline{42}}\times {\rm Spin} (0)\, .
    \end{eqnarray}  
\vskip0.2cm
\noindent
Hence in summary, the ${\cal N}=4$ massive super-(Weyl)$^2$ gravity theory contains $n_B+n_F=288$ degrees of freedom.

\subsection{Massless theory}

Now we need  the branching rules of the massive $USp(8)$ R-symmetry group
 into the R-symmetry group $SU(4)$ of the massless states. The specific decomposition of $USp(8)\rightarrow SU(4)$ for the relevant representations is as follows:
\begin{eqnarray}
 {\underline{8}}&=&      { {\underline{{ 4}}}}\oplus{\overline {\underline{{ 4}}}}    \, ,  \nonumber\\
  {\underline{27}}&=&      { {\underline{{ 6}}}}\oplus{\overline {\underline{{ 6}}}}  \oplus{ {\underline{{ 15}}}}  \, ,  \nonumber\\
{\underline{42}}&=& { {\underline{{ 1}}}}\oplus{\overline {\underline{{ 1}}}}\oplus { {\underline{{ 10}}}}+{{\underline{{ {\overline {10}}}}}}\oplus{ {\underline{{ 20'}}}}\, , \nonumber \\
{\underline{48}}&=&   { {\underline{{ 20}}}}\oplus{\overline {\underline{{ 20}}}}  \oplus    { {\underline{{ 4}}}}\oplus{\overline {\underline{{ 4}}}}  
\end{eqnarray}

\vskip0.4cm\noindent
Then the spectrum of the massless ${\cal N}=4$ Super-Weyl theory has the following form:
\vskip0.5cm
\noindent
(i) A  standard massless spin-two supergravity multiplet with $n_B+n_F=32$ degrees of freedom as given in eqs.(\ref{gn4a}) and (\ref{gn4b}).

\vskip0.2cm
\noindent
(ii)  In the non-standard sector,   we get first  from the massive Weyl multiplet  $w_{{\cal N}=4}$ a massless ghost-like spin-two supermultiplet  with $n_B+n_F=32$ and with the  helicites
and $SU(4)$ quantum numbers, again as given eqs.(\ref{gn4a}) and (\ref{gn4b}).

Second we get from $w_{{\cal N}=4}$ four massless spin-3/2 supermultiplets (in total $n_B+n_F=128$) with the following helicities  and $SU(4)$ representations, 
namely\footnote{The group theory of the massive AdS graviton Higgsing was discussed before in    \cite{Bachas:2017wva}.}
\begin{eqnarray}
{{\underline{\bar{ 4}}}}\times\lbrack   ({3\over 2},\underline 1) +  ( 1,\underline 4 )+({1\over 2},\underline 6)+
( 0 ,\bar{\underline 4})+(-{1\over 2},\underline 1)
\rbrack  \, ,
\end{eqnarray}
together with the CPT conjugate states
\begin{eqnarray}
{{\underline{{ 4}}}}\times\lbrack   ({1\over 2},\underline 1) +  ( 0,\underline 4 )+(-{1\over 2},\underline 6)+
( -1 ,\bar{\underline 4})+(-{3\over 2},\underline 1)
\rbrack 
  \, .
\end{eqnarray}
They contain the 15 gauge bosons of the local $SU(4)_R$ gauge symmetry.

 In addition, the massive Weyl multiplet  $w_{{\cal N}=4}$ contains six ${\cal N}=4$ vector multiplets of the form:
  \begin{equation}
 6~ ({\rm spin-one}):~  {{\underline{{  6}}}}\times\lbrack  
  (+1,\underline 1) +   (+{1\over2},\underline4 )
+( 0,\underline 6 )+(-{1\over2},\underline{\overline 4} )+(-1,\underline1).
\rbrack  
   \end{equation}
However these multiplets are unphysical since they can be gauged away by the superconformal transformations together with the local $SU(4)_R$ transformations.
Specifically, one of the 36 scalars in these vector multiplets is a Weyl mode.
Other  15 scalars are the helicity 
zero component of the massive vectors inside $w_{{\cal N}=4}$, which are gauged away by the local $SU(4)_R$ transformations.
Hence all
six vector-multiplets are unphysical, do not propagate and get removed from the spectrum.

We should note that the dipole ghost graviton and the tripole ghost 
spin-$3/2$  sector are accompanied by a dipole ghost complex scalar since the action is a higher-derivative action. Indeed,  the equations of motion  are fourth-order for the spin-2 and third order for the spin-$3/2$ states. 
This fact is also discussed in \cite{Johansson:2018ues} at the Lagrangian level. 
This is not the case for the $SU(4)$ gauge bosons which have standard Yang Mills
action. 
The sugra higher derivative action also contains a singlet vector mode
which, together with the gauge bosons, is part of the higher derivative
gravitino action (which as pointed out above obeys third order equations of motion). In other words,  the cubic gravitino action simultaneously describes
the gravitino, the partner of the graviton, as well as the gravitini of the gravitino multiplet. 

\vskip0.2cm
\noindent
Hence in summary, the massless ${\cal N}=4$ super-(Weyl)$^2$ gravity theory contains $n_B+n_F=192$ physical, propagating degrees of freedom.
The same spectrum  was also obtained in  \cite{Berkovits:2004jj} using the string twistor formalism for the construction of ${\cal N}=4$ super-(Weyl)$^2$ gravity.


\section{Conclusions and Outlook}

In this paper we obtained the spectrum of all  existing ${\cal N}$-extended Weyl$^2$ supergravities (${\cal N}=1,2,3,4$) in four space-time dimensions. 
We are summarizing the physical spectrum, ie. the number of super-multiplets of the massless theories after subtracting the gauge multiplets,
in the following table, where  we indicate in the first column the  top helicities
of each multiplet together with its CPT conjugate multiplet:

\vskip0.5cm
{
\tiny

\begin{equation}
\begin{tabular}{|c||c|c|c|c|}
\hline
   & ${\cal N}=4$                   & ${\cal N}=3$  & ${\cal N}=2$ & ${\cal N}=1$  \\
  \hline\hline
 $h_{\rm max}=2$  & 2 &        &             &  \\
 $h^{\rm CPT}_{\rm max} =0$  & 2 &        &         & \\
\hline
$h_{\rm max}=3/2$ & 4 & &  & \\
$h^{\rm CPT}_{\rm max}=1/2$ & 4&  &   &\\
\hline\hline
  $h_{\rm max}=2$  &  &   2     &             &  \\
 $h^{\rm CPT}_{\rm max} =-1/2$  &  &   2     &         & \\
\hline
$h_{\rm max}=3/2$ &  & 3&  & \\
$h^{\rm CPT}_{\rm max}=0$ & & 3 &   &\\
\hline
$h_{\rm max}=1$ &  & 1&  & \\
$h^{\rm CPT}_{\rm max}=1/2$ & & 1 &   &\\
\hline\hline
 $h_{\rm max}=2$ &  &        &  2          &  \\
 $h^{\rm CPT}_{\rm max} =-1$  &  &        &       2  & \\
\hline
$h_{\rm max}=3/2$ &  & &2  & \\
$h^{\rm CPT}_{\rm max}=-1/2$ & &  &  2 &\\
\hline
$h_{\rm max}=1$ &  & & 1 & \\
$h^{\rm CPT}_{\rm max}=0$ & & &  1 &\\
\hline\hline
 $h_{\rm max}=2$ &  &        &           & 2 \\
 $h^{\rm CPT}_{\rm max} =-3/2$  &  &        &         & 2\\
\hline
$h_{\rm max}=3/2$ &  & &  & 1\\
$h^{\rm CPT}_{\rm max}=-1$ & &  &   &1\\
\hline
$h_{\rm max}=1$ &  & &  &1 \\
$h^{\rm CPT}_{\rm max}=-1/2$ & & &   &1\\
\hline\hline
  \end{tabular}
\end{equation}
}

\vskip0.5cm

Note that in the supersymmetric Weyl$^2$ theories one can have bosonic terms with four derivatives 
and also some with only two derivatives.
This happens when the two-derivative terms would be auxiliary fields in Einstein supergravity.  In particular this is the case in $U({\cal N})$
Weyl$^2$ supergravity, where the Langrangian for the $U({\cal N})$ vector bosons is given by  the canonical $ F^{\mu\nu} F_{\mu\nu }$ term.\footnote{An
 analogous effect can be seen from the Wess-Zumino Langrangian 
\begin{equation}
{\cal L}=-{1\over 2}\partial^\mu z\square \partial_\mu \bar z -{i\over 2}\bar\psi\slashed{\partial}^3\psi+{1\over 2}\overline F\square F\, ,
\end{equation} 
which contains three Wess-Zumino multiplets: the first term is a dipole spin 0, the second  is a tripole ghost spin 1/2, whereas the last term describes two standard scalars.}
In addition we also discussed
some operators of higher dimension, corresponding to the  Bell-Robinson tensor, which is basically the square of the Weyl  tensor, or the square of the Cotton tensor.

Another issue we would like to point out here is that the (super)-conformal symmetry of the (super)-Weyl$^2$ theory is a classical symmetry. It is true that these theories are power-counting renormalizable  but the  one-loop beta-functions turns out to be non-vanishing \cite{FTs1} and therefore they suffer from a  conformal anomaly. 
The latter leads to  serious problems  in Weyl gravity as  the conformal symmetry  is gauged and therefore leads to  inconsistencies \cite{Duff,FTs2,Ts3}. 
Let us also note that one  may arrive at the same conclusion  by the calculation of the  chiral gauge anomalies of the $SU(4)$ R-symmetry \cite{RvN} and   
 by recalling that all anomalies 
are arranged in  the same   multiplet of the ${\cal N}=4$ superconformal symmetry.  

However, surprisingly, the ${\cal N}=4$ super-Weyl$^2$ gravity can be made UV-finite \cite{FTs2,FTs3}, and thus anomaly-free \cite{RvN} as well. For ${\cal N}=4$ Poincar\'e supergravity it has been conjectured in 
\cite{FKvP} a hidden superconformal symmetry.  This can be achieved by appropriate coupling the Weyl$^2$ supergravity to four  vector supermultiplets. 
Although this result has been shown to hold at one-loop level, it
is expected to hold  to all loops since the contributions to the $\beta$-function  and the conformal anomaly of SYM is only  an one-loop  effect for ${\cal N}>1$ conformal supergravities \cite{FTs3}. For the ${\cal N}=4$ case in particular, the above statement is strengthen by the fact that the conformal anomaly is connected by supersymmetry to the $SU(4)$ chiral anomaly which arises at one-loop also.

Let us notice  that  super-Weyl$^2$ gravity is not the only superconformal theory. 
As it is very well known, the ${\cal N}=4$ SYM  theory is also invariant under the rigid  superconformal supergroup $SU(2,2|4)$. Therefore, it can naturally source the ${\cal N}=4$ super-Weyl$^2$ gravity. In addition, the latter may affect the standard AdS/CFT holography. Indeed, allowing  a boundary 
Weyl$^2$ operator still preserves  the superconformal symmetry and it may correspond to a deformation of the holographic bulk theory.

Moreover, it could be, as pointed out in \cite{Ferrara:2018iko}
that   the pure ${\cal N}$-extended (Weyl)$^2$ supergravity theory in four dimensions is the holographically dual
boundary theory
of an $AdS_5$ bulk theory, which is a higher spin theory with a spin-four multiplet of the $2{\cal N}$-extended supersymmetry algebra in five dimensions.
These kind of theories,  denoted by W-supergravities, were recently constructed \cite{Ferrara:2018iko} in flat four-dimensional space-time using a double copy and S-fold construction.
 Within such a duality,  one would expect that a conformal   operator respectively higher tensor   $ T_{\mu\nu\rho\sigma}$ on the boundary,
which is quadratic in  the curvature tensors and their derivatives, and which is also divergence-free on-shell,
is coupled to a spin-four field in the bulk.
 It would be interesting to find  the exact 5D holographic dual, if any, of the 4D Weyl supergravity.

Alternatively, one may consider super-Weyl$^2$ gravity as a bulk theory. In that case, $AdS_4$ is still a vacuum solution of the theory and there should exists a holographic 3D theory in the boundary, which   it might not be  unitary.  Indeed,  the massless graviton will still be coupled to a conserved spin-2 operator (energy-momentum tensor) at the boundary, the massless vector will be coupled to a conserved spin-1 operator, whereas it is expected that the massive spin-2 ghost to couple to a non-unitary spin-2 operator.   Furthermore,  it would be very interesting to see if there is a corresponding string construction of   holographic duality, possibly
using the twistor string approach of \cite{Berkovits:2004jj} for the construction of the superconformal Weyl$^2$ theory.

\section*{Acknowledgements:}

We gratefully acknowledge correspondence with N. Boulanger and F.
Farakos.
 The work of S.F. is supported in part
by CERN TH Dept and INFN-CSN4-GSS. The work of D.L. is supported by the ERC Advanced Grant ``Strings and Gravity" (Grant No. 320045) and the Excellence Cluster Universe.
He also is grateful to the CERN theory department for its hospitality, when part of this work was performed. 
A.K. is supported by the GSRT under the EDEIL/67108600.

\end{document}